\documentclass[twocolumn]{article}

\usepackage{lipsum}
\usepackage{todonotes}
\usepackage{amsmath}
\usepackage{placeins}
\usepackage{booktabs}
\usepackage{color}
\usepackage{hyperref}
\usepackage[margin=1.8cm]{geometry}		 			
\usepackage[font=small,labelfont=bf]{caption}		
\usepackage{enumitem}

\usepackage{amssymb}
\usepackage{mathtools}
\DeclarePairedDelimiter{\norm}{\lVert}{\rVert}

\usepackage{enumitem}
\usepackage{authblk}

\graphicspath{Figures}

\newcommand{\head}[1]{\textbf{\emph{}}}
\newcommand{\caphead}[1]{\textbf{#1}}

\newcommand{\revA}[1]{{#1}}   						
\newcommand{\revB}[1]{{#1}}   						
\newcommand{\revC}[1]{{#1}}							
\newcommand{\rev}[1]{#1}			

\newcommand{\chapternote}[1]{{%
		\let\thempfn\relax
		\footnotetext[0]{\emph{#1}}
}}

\usepackage{etoolbox}

\title{Gaussian process regression for forecasting battery state of health}
\author[1]{Robert R. Richardson}
\author[1]{Michael A. Osborne}
\author[1]{David A. Howey}
\affil[1]{Department of Engineering Science, University of Oxford, Oxford, UK}

\begin{document}

\twocolumn[
\begin{@twocolumnfalse} 
	\maketitle
	\begin{abstract}
		\textbf{Accurately predicting the future capacity and remaining useful life of batteries is necessary to ensure reliable system operation and \revA{to} minimise maintenance costs.
			The complex nature of battery degradation has meant that mechanistic modelling of capacity fade has thus far remained intractable;
			however, with the advent of cloud-connected devices, data from cells in various applications is becoming increasingly available, and the feasibility of data-driven methods for battery prognostics is increasing.
			Here we propose Gaussian process (GP) regression for forecasting battery state of health, and highlight various advantages of GPs over other data-driven and mechanistic approaches.
			GPs are a type of Bayesian non-parametric method, and hence can model \revA{complex} systems whilst handling uncertainty in a principled manner.
			Prior information can be exploited by GPs in a variety of ways: explicit \revC{mean} functions can be used if the functional form of the underlying degradation model is available, and multiple-output GPs can effectively exploit correlations between data from different cells.
			We demonstrate the predictive capability of GPs for short-term and long-term (remaining useful life) forecasting on a selection of capacity vs.\ cycle datasets from lithium-ion cells.
		}
		\vspace{4mm}
	\end{abstract}
\end{@twocolumnfalse}
]




\subsection*{Keywords}
Lithium-ion battery, Gaussian process regression, State-of-Health, degradation, prognostics
\chapternote{\textsuperscript{1}Author contact information: robert.richardson@eng.ox.ac.uk, mosb@robots.ox.ac.uk, david.howey@eng.ox.ac.uk.}

\subsection*{Highlights}
\begin{itemize}
	\item Gaussian process (GP) regression used for forecasting battery state of health
	\item Systematic kernel function selection allows fitting of complex degradation behaviour
	\item Explicit mean functions combine GPs with knowledge of cell degradation mechanisms
	\item Multi-output GPs effectively exploit correlations between data from different cells
\end{itemize}
\begin{figure}[h]
	\centering
	{\includegraphics[width=1\columnwidth]{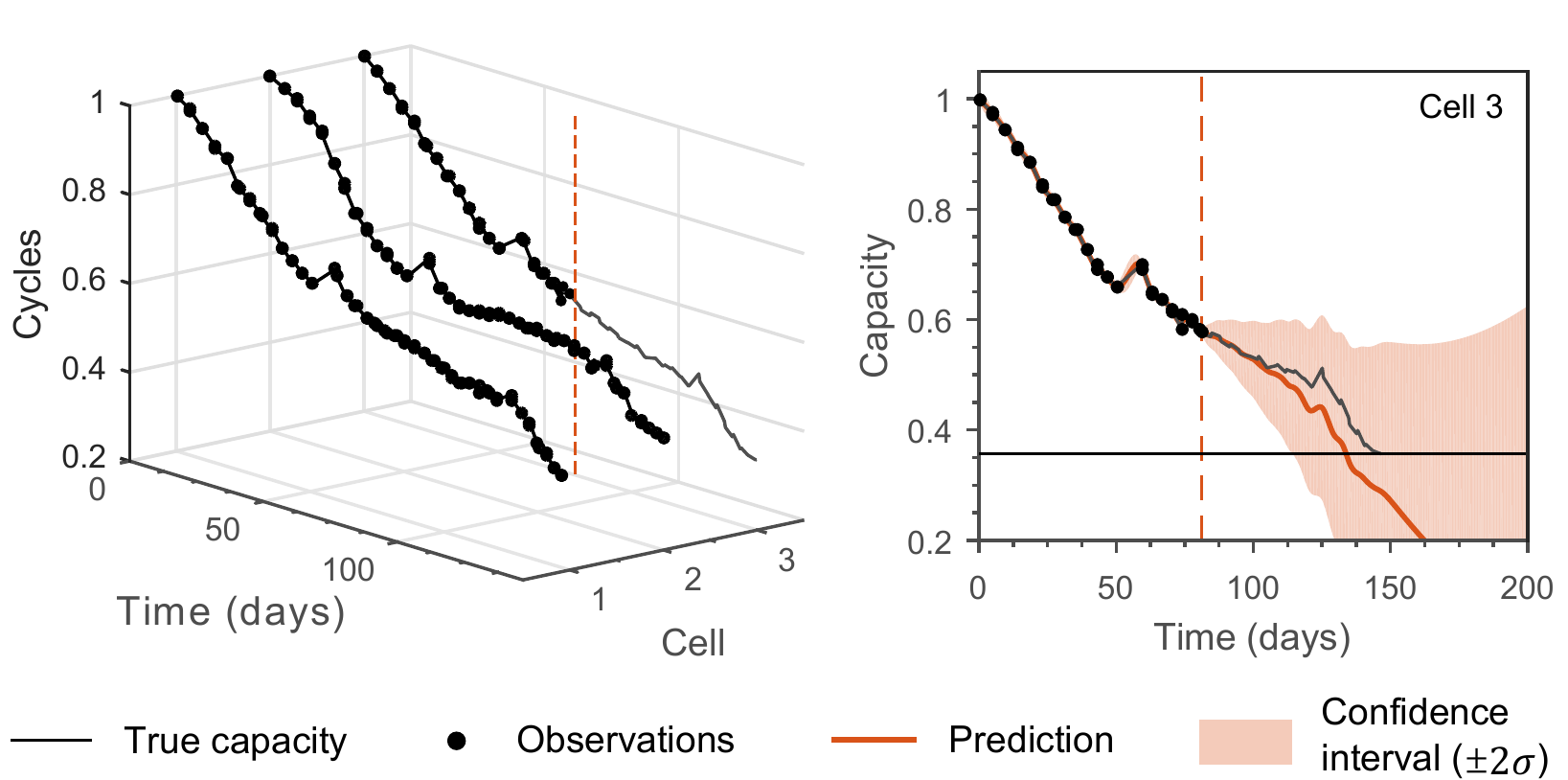}}
	\caption{Graphical abstract}
\end{figure}

	\section{Introduction}
	
	\head{Motivation}
	Lithium-ion batteries (LIBs) are increasingly playing a pivotal role in applications ranging from transport to grid energy storage.
	\rev{However, not knowing a battery's rate of capacity loss or useful life renders the system susceptible to an unanticipated decline in performance or to operate in an unsafe regime~\cite{cabrera2016calculation}.
	To mitigate this, LIBs are often over-sized and under-used, which results in unnecessary cost inefficiencies.
	Second life applications -- which offer a potential means of offsetting high initial battery costs in EV applications~\cite{neubauer2011ability,birkl2014modular} -- rely particularly heavily on accurate capacity forecasting, since this determines the potential value of a cell in its secondary application.
	Hence, accurate prognostics is an important component of a modern battery management system.}
	
	\head{Battery degradation}
	Since the performance capability of a cell is largely defined by its nominal capacity and internal resistance, the State of Health (SoH) is typically defined by one or both of these parameters.
	In the present case we focus on capacity estimation, although the methods we employ could be applied in either case.
	Predicting the future state of a LIB is non-trivial due to the complex interplay of parameters and the path-dependence of the degradation behaviour\revC{~\cite{birkl2017degradation}}.
	
	\head{Conventional approaches}
	The conventional approach to SoH forecasting relies on degradation modelling via electrochemical or equivalent circuit models.
	Electrochemical models enable some physical interpretation of degradation behaviour; however, simulating all the underlying dynamics responsible for battery degradation is a momentous challenge.
	Semi-empirical models have also been used to capture the dependence of battery SoH on likely stressing factors. For instance, Ref.~\cite{wang2011cycle} develops a capacity fade model using temperature, depth-of-discharge (DoD), C-rate and time as inputs.
	These models have had some success, although their accuracy is limited when environmental and load conditions differ from the training data set, and when the capacity fade depends on additional contributions from unknown sources.
	
	\head{Data-driven approaches}
	Data-driven approaches are gaining attention due to the increasing availability of large quantities of battery data.
	\revC{There }are various ways data-driven techniques could be applied, and each amounts to different assumptions about the nature of the underlying processes.
	The most common and \revC{simple }of these is to use a direct mapping from \revC{cycle }to SoH~\cite{goebel2008prognostics,saha2008uncertainty,he2011prognostics}.
	Simplistically, this amounts to fitting a curve to the capacity-cycle data, and then predicting future values by extrapolating the fitted curve.
	\revC{This implies that accurate capacity data for some previous cycles in the battery life is available.
	We note that battery capacity estimation is another important topic; however, the primary concern of this paper is capacity \emph{forecasting}, i.e. estimating future values of capacity. Hence, we assume that capacity-cycle data is available\revA{ -- }in practice such data may be acquired by direct measurement (slow-rate charge-discharge cycles specifically applied at periodic intervals for capacity measurement) or by a variety of other techniques which obviate the need to interfere with the system (such as parameter estimation of equivalent circuit models). For a detailed review of methods for capacity estimation see Ref.~\cite{farmann2015critical}.}
	
	\begin{figure*}[hbt]
		\centering
		{\includegraphics[width=0.9\textwidth]{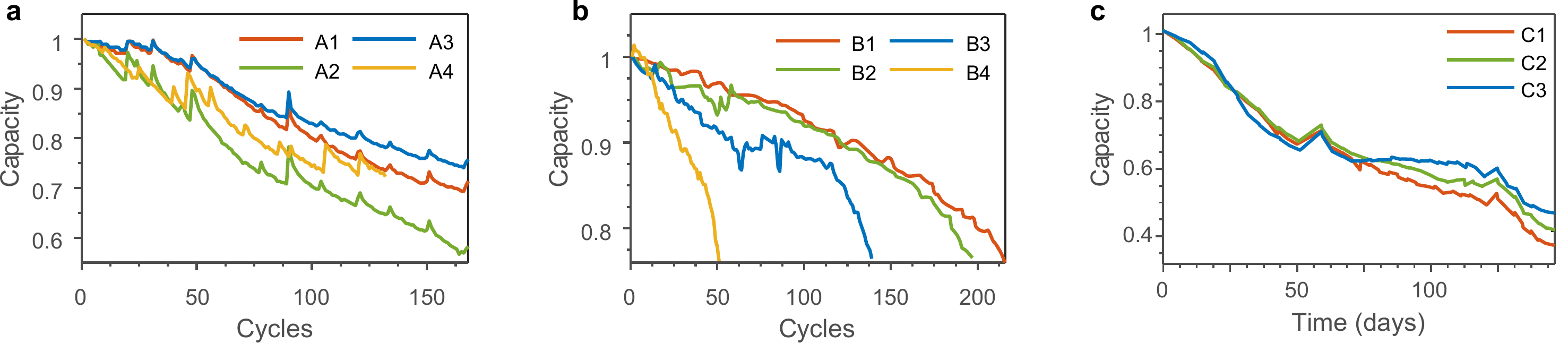}}
		\caption{\caphead{Battery datasets.} \textbf{a}, NASA Battery Data Set used with the basic, single-output GP (Figs.~\ref{fig:Figure_gpKernelSearch} - \ref{fig:Figure_gpNaiveRuL}), \textbf{b}, Data extracted from Liu et al.~\cite{liu2013prognostics} and used with the explicit mean function GP (Fig.~\ref{fig:Figure_gpExplicitMean}) \textbf{c}, NASA Randomized Battery Usage Data Set used with the multi-output GP (Fig.~\ref{fig:Figure_gpMulti1}). \revA{Note that the capacity is normalised against the starting capacity in each case}.}
		\label{fig:Figure_gpData}
	\end{figure*}	

	\head{Mapping from cycle to SoH}
	On the one hand, mapping from cycle to SoH is over-simplistic since the cell capacity depends on various factors, and the historical capacity data alone is unlikely to be sufficient for predicting future capacity.
	On the other hand, it is reasonable to expect the previous capacity to be \emph{somewhat} correlated with future capacity and hence it is worth exploring the limits of its predictive capability.
	Moreover, the methods applied to capacity vs.\ cycle data could subsequently be applied to more informative (possibly higher dimensional) inputs (such as estimates of physical parameters such as lithium inventory or active material vs.\ cycle; see \cite{birkl2017degradation}).
	
	\head{Non-parametric modelling}
	\revA{A key advantage of our approach is that it is non-parametric. Non-parametric methods permit a model expressivity (e.g a number of parameters) that is naturally calibrated to the requirements of the data.} 
	Hence, such methods can model arbitrarily complex systems, provided enough data is available.
	For instance, a number of recent studies have used Support Vector Machines \revC{(SVMs)} for predicting future cell capacity based on it's historical capacity vs.\ cycle data and/or from data from multiple identical cells~\cite{hu2016battery,patil2015novel,wang2013prognostics,nuhic2013health}.
	The success of these works demonstrates the advantage of \revC{such} approaches.

	\head{Uncertainty management}
	However, an important aspect of prognostics is not only predicting future values of the variable of interest but also expressing the uncertainty associated with these values.
	Bayesian methods provide a principled approach to dealing with uncertainty.
	This results in a credible interval comprising \revA{probabilistic} upper and lower bounds, which is essential for making informed decisions.
	GPs are a non-parametric Bayesian method \revC{that offer a number of other unique advantages which have not been fully exploited in prior work.}

	\head{Previous GP studies}
	There have been a limited number of studies investigating GPs for battery prognostics. Goebel et al.~\cite{goebel2008prognostics} investigated the use of GPs for extrapolating battery internal resistance and subsequently deriving capacity estimates based on a linear relationship between resistance and capacity.
	They showed that GPs could handle the non-linear data manifested by battery degradation but they concluded that although \revC{they were} capable of characterizing the uncertainty in the predictions, \revC{they} lacked long-range predictive capability.
	Recently, Liu et al.~\cite{liu2013prognostics} applied Gaussian process regression to battery capacity prediction, and showed that \revC{their} predictive accuracy was improved when a linear or quadratic Explicit \revA{Mean} Function (EMF; see Section~\ref{sec:methods-EMFs}) was used. However, the assumption of a linear or quadratic function for the underlying battery behaviour is overly simplistic; it would be preferable to use mean functions inspired by battery degradation models.
	In a separate study by the same authors~\cite{li2015new}, a Mixture of Gaussian Processes model was used to initialise the parameters of a parametric model using data from identical cells. The model parameters were then recursively updated using a particle filter. Whilst this made use of data from multiple cells, it merely used the data as a means of initialising a parametric model. Superior performance can be achieved by using multi-output models to capture correlations between the capacity trends in each cell, as we show in the present work.

	\head{Contributions of this paper}
	\revC{Existing studies fail to exploit many of the capabilities of GPs -- in the current study, we present a thorough analysis of these capabilities.}
	Specifically, we use GPs for short-term and long-term, \revC{i.e.\ remaining useful life (RUL),} forecasting on a selection of capacity vs.\ cycle datasets from lithium-ion cells (Fig.~\ref{fig:Figure_gpData}).
	First, the most basic GP is studied. We highlight the importance of systematically selecting the correct kernel function (an issue which has been overlooked in previous works) and the advantages of using compound kernel functions.
	We then present two extensions to this basic approach which enable improved performance: (i) we use explicit mean functions based on known parametric battery degradation models to exploit prior knowledge of battery degradation behaviour and (ii) we use multi-output GPs to effectively exploit available capacity data from multiple identical cells.
	Lastly, it is worth underscoring the fact that all the methods presented here are rigorously evaluated using different proportions of training data (i.e. using capacity data up to the current cycle for training, with various different values of the current cycle).
	This is in contrast to most previous studies on battery prognostics, which merely evaluate the accuracy of the predictions made at a single arbitrarily selected cycle (e.g. the first half of training data).

\section{Methods}
	
	The goal of a regression problem is to learn the mapping from \revA{inputs ${x}$ to outputs ${y}$}, given a labelled training set of input-output pairs $\mathcal{D} = \{(x_i, y_i)\}_{i=1}^{N_D}$, where $N_D$ is the number of training examples. In our case, \revA{the input ${x}_i \in \mathbb{Z}^{+}$ is} the integer number of cycles applied up to the current cycle, and \revA{the output ${y}_i \in \mathbb{R}^{+}$ is} the corresponding measured capacity (\revC{all }capacities are normalised against the initial, maximum capacity).
	We assume the underlying model takes the form \revA{${y} = f({x}) + \revB{\varepsilon}$, where $f({x})$} represents a latent function and $\revB{\varepsilon} \sim \mathcal{N}(0, \sigma^2)$ is \revA{an independent and identically distributed noise contribution}.

	The learned model can then be used to make predictions at test indices $\mathbf{x^*} = \{x_i^*\}_{i=1}^{N_T}$ (cycles at which we wish to estimate the capacity) for unknown observations $\mathbf{y^*} = \{y_i^*\}_{i=1}^{N_T}$, where $N_T$ is the number of test indices.
	In our case we are interested in \revC{extrapolation to forecast }future values \revC{of }capacity \revC{(}and so the test indices are the future \revA{cycles} up until the \revC{end of life} (EoL)\revC{)}. The EoL is reached when the capacity drops below a predefined threshold denoted by $y_{\text{EoL}}$; the corresponding cycle number at which this occurs is denoted $x_{\text{EoL}}$.
	\revA{Note that $x_{\text{EoL}}$ is a-priori unknown, we will infer it using our model.}

	We evaluate our methods using two different metrics, which reflect the quantities of interest in a practical application\revC{: the first is} the root-mean-squared error (RMSE) in the capacity estimation, which we denote $\text{RMSE}_Q$. At a given cycle, $c$, where we train using data up to the current cycle ($\mathbf{x} = [1,2,\ldots\, c]^T$) and test on the remainder ($\mathbf{x}^* = [c{+}1, c{+}2,\ldots\, x_{\text{EoL}}]^T$), $\text{RMSE}_Q$ is defined as 
	\begin{equation}
	\text{RMSE}_Q(\hat{y_i}^*, y_i^*) = \sqrt{\frac{1}{N_T}\sum_{i=1}^{N_T}\left(\hat{y_i}^* - y_i^*\right)^2}
	\end{equation}
	\revA{where $\hat{y}_i$ is the estimate given data only up to and including $c$.}
	By taking only one training proportion (i.e. a single value of $c$), we would obtain just a single $\text{RMSE}_Q$ value. This would run the risk of misrepresenting the true performance of the method over the full cycle-life.
	For instance, it would not be acceptable for the estimates to be accurate after the first 30 cycles are observed but to then diverge when the next 5 cycles are received.
	Hence, in order to thoroughly validate our methods, we test the performance using all values of $c$ from 20\% of the cycle-life onwards.
	We thus obtain a value of $\text{RMSE}_{Q}$ at each cycle, and can plot $\text{RMSE}_{Q}$ vs.\  cycle (see later).
	\revC{This }is \revA{in} contrast to most previous studies, which use just a single arbitrary value of $c$.

	The second \revC{metric is} the RMSE in the EoL prediction, which we denote $\text{RMSE}_{EoL}$.
	\revA{For each value of $c$}, there is a single EoL prediction. Hence, we can compare the predictions at all $c$ values against the true EoL to obtain a mean error, defined as
	\begin{equation}
	\text{RMSE}_{\text{EoL}}(\revA{\hat{x}_{\text{EoL}, \, j}^*}, x_{\text{EoL}}^*) = \sqrt{\frac{1}{N_c}\sum_{j=1}^{N_c}\left(\revA{\hat{x}_{\text{EoL}, \, j}^*} - x_{\text{EoL}}^*\right)^2}
	\end{equation}
	where \revA{$\hat{x}_{\text{EoL}}$ is the estimate given data only up to and including $c$, and} \revA{$N_c$} is the number of cycles at which we test \revA{(i.e. the number of different values of $c$)}.

	Note that this metric neglects the intermediate values of the capacity between the current cycle and the EoL. For instance, two cells could have very different capacity trajectories over the duration of their lives, whilst still reaching their EoL after a similar number of cycles. Hence, a good model should have low values of both of the above metrics.

	\subsection{Gaussian process regression}
		
	This section gives an overview of Gaussian process regression.
	For simplicity, our presentation assumes the inputs and outputs are scalar, since we only consider 1-D capacity vs.\ cycle data in this work. However, the analysis can easily be extended to multidimensional inputs, if desired.
	A more detailed presentation of \revA{Gaussian process regression (GPR)} is given in Chapter~15 of~\cite{murphy2012machine}, and a more comprehensive book on the topic is~\cite{rasmussen2006gaussian}.
	
	A Gaussian process (GP) defines a probability distribution over functions, and is denoted as:
	\begin{equation}
		f({x}) \sim \mathcal{GP}\bigl(m({x}), \kappa({x},{x}')\bigr),
	\end{equation}
	where $m({x})$ and $\kappa({x},{x}')$ are the mean and covariance functions respectively, denoted by
	\begin{align}
		m({x}) & =
		\mathbb{E} [f({x})], \\
		\kappa({x}, {x}') & =
		\mathbb{E} [\left(f({x}) - m({x})\right) \left(f({x}') - m({x}')\right)^T].
	\end{align}
	For any finite collection of input points, say $\mathbf{x} = {x}_1,...,{x}_{N_D}$, this process defines a probability distribution $p\left( f({x}_1),...,f({x}_{N_D}) \right)$ that is jointly Gaussian, with some \revA{mean $\mathbf{m} (\mathbf{x})$ and covariance $\mathbf{K} (\mathbf{x})$ given by $K_{ij} = \kappa({x}_i,{x}_j)$.}
	
	Gaussian process regression is a way to \revC{undertake} non-parametric regression with Gaussian processes.
	The key idea is that, rather than postulating a parametric form for the function $f({x, \theta})$ and estimating the parameters $\theta$ (as in parametric regression), we instead assume that the function $f({x})$ is a sample from a Gaussian process as defined \revA{above.}
	
	The most common choice of covariance function is the squared exponential (SE), defined by
	\begin{equation}
		\kappa_{\text{SE}}({x}, {x}') = \theta_f^2 \exp\left(-\frac{1}{\theta_l^2} \norm{{x}-{x}'}^2 \right)
		.
	\end{equation}
	The covariance function parameters\footnote{The term `non-parametric' is evidently a misnomer since the covariance function contains parameters; however, these are technically hyperparameters~\cite{rasmussen2006gaussian} since they are the parameters of a \emph{prior} function.}, $\theta_f$ and $\theta_l$, control the $y$-scaling and $x$-scaling, respectively.
	
	The SE kernel is a \emph{stationary} kernel, since the correlation between points is
	purely a function of the difference in their inputs, ${x}-{x}'$. We only consider stationary kernels in this work.
	\revA{The choice of the SE kernel makes the assumption that the function} is very smooth (infinitely differentiable).
	This may be too strict a condition for many physical phenomena~\cite{stein2012interpolation}, and so a common alternative is the Mat\'ern covariance class:
	\begin{multline}
		\kappa_{\text{Ma}}({x}-{x}') = \\
		\sigma^2 \frac{2^{1-\nu}}{\Gamma(\nu)}
		\left(\sqrt{2\nu}\frac{({x}-{x}')}{\rho}\right)^{\nu}
		\mathcal{R}_{\nu}\left(\sqrt{2\nu}\frac{({x}-{x}')}{\rho}\right)
		,
	\end{multline}
	where $\nu$ is a smoothness hyperparameter (larger $\nu$ implies smoother functions)
	and $\mathcal{R}_{\nu}$ is the modified Bessel function. This equation simplifies considerably for half-integer $\nu$. The most common examples are $\nu = 5/2$ and $\nu = 3/2$, which we denote as Ma5 and Ma3 in this work.
	The final covariance we \revA{consider} in this paper is the periodic covariance,
	\begin{equation}
		\kappa_{\text{Pe}}({x},{x}') = \theta_f^2 \exp\left( -\frac{2}{\theta_l^2} \sin^2 \left(\pi\frac{ {x}-{x}'}{p} \right) \right)
	\end{equation}
	which is suitable for functions with periodic behaviour. \revA{The hyperparameter $p$ is the period of $f(x)$.}
	
	Compound kernels can be created by affine transformations of individual kernels.
	\revA{We} limit our attention in this paper to \revA{addition of} kernels, since these were found to be capable of expressing the structure of the battery data under study, and since they lead to greater ease of interpretation than multiplicative kernels. Ref.~\cite{duvenaud2013structure} provides a more detailed discussion of kernel composition and also addresses the issue of automating the choice of kernels.
	In summing kernels, the \revC{data are} modelled as a superposition of independent functions. This can be interpreted as different processes operating at different input and/or output scales.
		
	The mean function is commonly defined as $m(\mathbf{x}) = 0$ since the GP is flexible enough to model the true mean arbitrarily well. In Section~\ref{sec:methods-EMFs}, we consider parametric models (based on battery degradation models) for the mean function, such that the GP models \revC{only }the residual errors.
	
	
	Now, if we observe a labelled training set of input-output pairs
	$\mathcal{D} = \{({x}_i, {y}_i)\}_{i=1}^{N_D}$, predictions can be made at test indices $\mathbf{x}^*$ by computing the conditional distribution $p(\mathbf{y}^* \vert \mathbf{x}^*, \mathbf{x}, \mathbf{y})$.
	This can be obtained analytically by the standard rules for conditioning  Gaussians~\cite{murphy2012machine}, and (assuming a zero mean for notational simplicity) results in a Gaussian distribution given by:
	\begin{equation}
		p(\mathbf{y}^* \vert \mathbf{x}^*, \mathbf{x}, \mathbf{y}) = \mathcal{N}(\mathbf{y}^* \vert \mathbf{m}^*, \mathbf{\Sigma}^*)
	\end{equation}
	where
	\begin{align}
		\mathbf{m}^* & = \mathbf{K}(\mathbf{x}, \mathbf{x}^*)^T
		\mathbf{K}(\mathbf{x}, \mathbf{x}^*)^{-1}
		\mathbf{y}\\
		\mathbf{\Sigma}^* & = \mathbf{K}(\mathbf{x}^*,\mathbf{x}^*)
		- \mathbf{K}(\mathbf{x}, \mathbf{x}^*)^T
		\mathbf{K}(\mathbf{x}, \mathbf{x}^*)^{-1}
		\mathbf{K}(\mathbf{x}, \mathbf{x}^*).
	\end{align}

	The values of the hyperparameters $\theta$ may be optimised by minimising the negative log marginal likelihood \revA{defined as $\text{NLML} = -\log p(\mathbf{y} \vert \mathbf{x}, \theta)$.}
	The NLML automatically performs a trade-off between bias and variance, and hence avoids over-fitting the data.	
	Given an expression for the NLML and its derivative w.r.t $\theta$ (both of which can be obtained in closed form), we can estimate $\theta$ using any standard gradient-based optimizer. In our case, we used the GPML toolbox~\cite{rasmussen2010gpml} implementation of \revA{conjugate gradients}. Since the objective is not convex, local minima can be a problem. However, this was not an issue in the present study, as was verified by repeated \revA{diverse} initialisations using Latin hypercube sampling~\cite{iman2008latin} yielding identical results.
	Minimising the NLML further allows us to perform model selection, i.e. to choose the kernel function, not just the values of the hyperparameters for a given kernel function. Kernel function selection is perhaps the most important aspect of GP modelling, yet it has not been addressed in a principled manner in the aforementioned \revA{battery degradation literature \cite{goebel2008prognostics,liu2013prognostics,li2015new}}

	\subsection{Explicit mean functions \label{sec:methods-EMFs}}
	
	Explicit mean functions (EMFs), also referred to as explicit basis functions \cite{rasmussen2006gaussian} or semi-parametric Gaussian processes~\cite{murphy2012machine}, allow us to express prior information we may have about the expected functional form of the model.
	For instance, let's say we have a battery degradation model which predicts capacity fade of the form \revB{$y = m({x}; \mathbf{\theta}_{\text{deg}})$} where $\mathbf{\theta}_{\text{deg}}$ are the parameters of the degradation model, but we believe that there may be other contributions to the battery capacity fade that the model does not account for.
	We can then model the capacity as the sum of a GP and the parametric model:
	\begin{equation}
		\revB{{y} = m({x}, \theta_{\text{deg}}) + f({x}, \theta) + \revB{\varepsilon}}
	\end{equation}
	This formulation expresses that the \revC{data are} close to the degradation model with the residuals being modelled by a GP (and a noise term).
	When fitting this model, we optimize over the degradation model parameters $\theta_{\text{deg}}$ jointly with the hyperparameters $\theta$ of the covariance function.

	\subsection{Multi-output GPs\label{sec:methods-multi-output}}

	If we have capacity vs.\ cycle data for multiple batteries undergoing similar loading profiles, we may expect the capacity trends to be correlated.
	This prior assumption can be modelled using multi-output GPs.
	This section draws largely from previous works on multi-output GPs \cite{osborne2010bayesian,durichen2015multitask}; and further details of similar methods can be found in those works.
	
	A function with multiple outputs can be dealt with by treating it as having a single output and an additional input.
	This additional (discrete) input, $l$, can be thought of as a label for the associated
	output.
	Let's say we have $m$ cells whose inputs and outputs are $\{\mathbf{x}^l,  \mathbf{y}^l\}_{l = 1}^{m}$, where $\{\mathbf{x}^l, \mathbf{y}^l\} = \{(x_i, y_i)\}_{i=1}^{N_l}$, and $N_l$ is the number of training points associated with cell $l$.
	Each input of the multi-output model is then a $1\times2$ vector defined as $\mathbf{x}_{i,l} = [\mathbf{x}^l(i), l]$, and this has an associated scalar output $y_{i,l} = \mathbf{y}^l(i)$.
	Assuming, for notational simplicity, \revB{that all cells are observed at the same set of cycles and hence} $N_l = n$ for all $l$, we can now write the entire set of inputs and outputs as $\mathbf{X} = \left\{\{\mathbf{x}_{i,j}\}_{i=1}^{n}\right\}_{l=1}^m$ and $\mathbf{y}=\left\{\{{y}_{i,j}\}_{i=1}^{n}\right\}_{l=1}^m$.
	A new covariance function can then be defined as the product of a label covariance and a standard covariance
	\begin{equation}
		\kappa_{\text{MOGP}}(x,x',l,l') = \kappa_l(l,l') \times \kappa_{{x}}({x}, {x}')
	\end{equation}
	where $\kappa_l$ captures the correlation between outputs, and $\kappa_{{x}}$ is the covariance with respect to cycles for a given output.
	The covariance matrix for all n cells is then the $mn \times mn$ matrix defined by
	\begin{equation}
		\mathbf{K}_{\text{MOGP}}(\mathbf{X}, \mathbf{L}, \theta_l, \theta_{\mathbf{x}}) = \mathbf{K}_l(l,\theta_l) \otimes \mathbf{K}_{\mathbf{x}}(\mathbf{X}, \theta_{x})
	\end{equation}
	where $\otimes$ is the Kronecker product, $\mathbf{L} = \{l\}_{j=1}^m$, and $\theta_l$ and $\theta_{{x}}$ are the hyperparameters for $\mathbf{K}_l$ and $\mathbf{K}_{x}$ respectively.
	Note that the assumption that $N_l=n$ for all cells can easily be relaxed \revB{such that the model may be applied to problems with capacities observed at different cycles for each cell}.
	
	Lastly, we parametrise the label covariance matrix using a spherical parametrisation scheme~\cite{osborne2010bayesian}:
	\begin{equation}
		\mathbf{K}_l = S^TS \text{diag} (\tau)
		\label{eq:spherical-parametrisation}
	\end{equation}
	where 
	$\tau = \{\tau_l\}_{l=1}^m$ is a vector of output scales corresponding to the different values of $l$,
	and $S$ is an upper triangular matrix of size $m \times m$, whose $l$th column contains the spherical coordinates
	in $\mathcal{R}^l$ of a point on the hypersphere $\mathcal{S}^{l-1}$, followed by the requisite number of zeros. For example, $S$ for a three dimensional space is
	\begin{equation}
		S =
		\begin{bmatrix}
		1 & \cos(\phi_1) & \cos(\phi_1) \\
		0 & \sin(\phi_1) & \sin(\phi_2)\cos(\phi_3) \\
		0 & 0 & \sin(\phi_2)\sin(\phi_3)
		\end{bmatrix}
	\end{equation}
	Note that this ensures that $S^TS$ has ones across its diagonal and hence all other entries may be thought of as akin to correlation coefficients, lying between $-1$ and $1$.
	
	The full set of hyperparameters\revC{, $\theta_{l}$,} therefore \revC{consists of} the values of $\phi_l$ and $\tau_l$. If the output scales are expected to be the same for each cell (as we assume in the present work), then the values of $\tau_l$ can be fixed to a single value, $\tau$.
	In this case, the total number of hyperparameters required is $\frac{1}{2}(m + 1)$.

	Once a suitable covariance has been defined, parametrisation and test prediction can be achieved in the same manner as that of single-output GPs, using \revA{optimisation of the NLML to select} the hyperparameters of the joint covariance matrix.
	The advantage of the multi-output scheme is that similarities between cells can be captured (through the shared hyperparameters, $\theta_x$), without imposing strict equivalence (through the cell-specific differences induced by the label hyperparameter, $\theta_l$).
	We employ a standard implementation of this method, which scales as $O(m^3 n^3)$.
	However, we note various different efficient/approximate schemes have been proposed in the literature, based on approximating input data with pseudo points~\cite{alvarez2011computationally}, exploiting grid structure~\cite{wilson2015kernel} and exploiting the recursivity of the estimation problem in online settings~\cite{osborne2010bayesian,pillonetto2010bayesian,samo2015p}; and these could be used if larger numbers of cells/training points were required.

\section{Basic single-output GP -- Results \label{sec:basic-gp}}

	\head{Overview}
	The first example we consider is a basic single-output GP with a constant prior mean set equal to the mean of the observed capacity data.
	The dataset considered consists of capacity vs.\ cycle data obtained from the NASA battery data repository (see Appendix~\ref{sec:appendix-data}, and Fig.~\ref{fig:Figure_gpData}a). Here, we present the results for Cell~A1, although similar results were obtained for Cells A2-A4.
	
	\subsection{Kernel function selection}
	\head{Short time-scale variation}
	It is apparent from Fig.~\ref{fig:Figure_gpData}a that the capacities experience a long-term downward trend with occasional, apparently discontinuous, step increases. In other words, \revC{they exhibit} a combination of short- and long- term structure.
	The physical explanation for the short term jumps is not clear -- it may in fact be an artefact of the measurement process.
	\rev{For instance, the data indicates that these increases tend to occur after long periods without cycling, possibly when reference tests were performed, which indicates that the capacity increase may be related to these pauses.}
	In any case, accounting for these local variations by means of appropriate kernel function selection is essential since: (i) the capacity measurement provided in a real application could also manifest similar artefactual variations, and (ii) accounting for such artefacts is necessary in order to correctly express the uncertainty in subsequent measurements obtained via the same process.
	
\begin{figure}[h]
		\centering
		\includegraphics[width=0.48\textwidth]{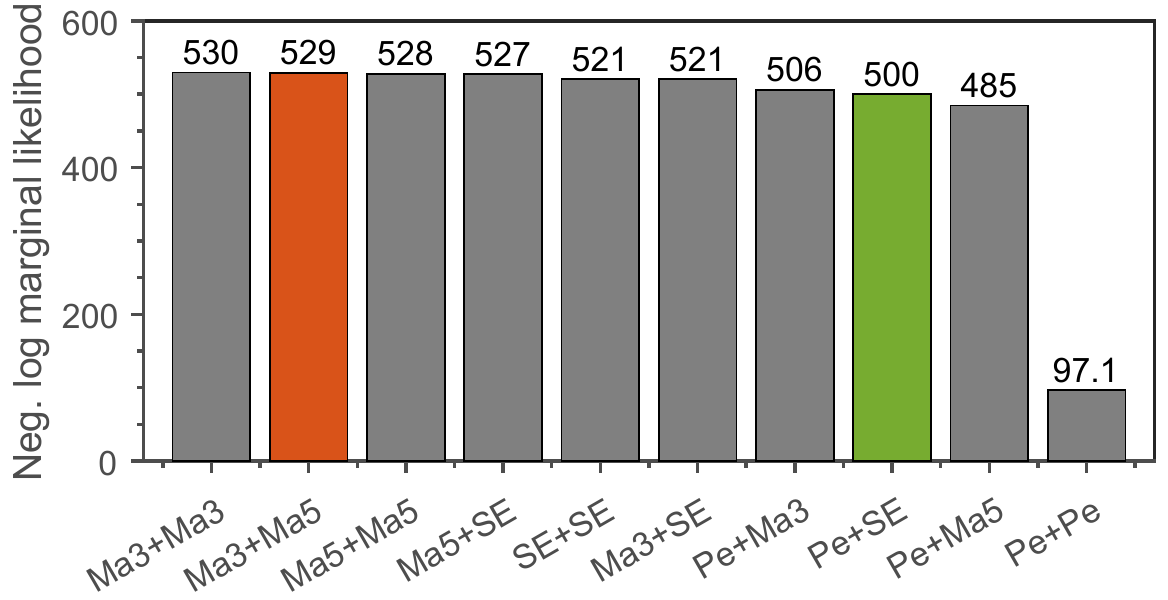}
		\caption{\caphead{Ranking of kernel function combinations by marginal likelihood.} The coloured bars for selected kernel pairs correspond to the coloured lines in Figs.~\ref{fig:Figure_gpDecomposeCov}a and \ref{fig:Figure_gpDecomposeCov}b.}
		\label{fig:Figure_gpKernelSearch}
	\end{figure}

	
	\head{Kernel function evaluation}
	In order to identify a suitable compound kernel function, we assessed 10 different compound kernels: namely, all possible pairs of the following base kernels:  \revC{Mat\'ern} 5/2 (Ma5), \revC{Mat\'ern} 3/2 (Ma3), Squared Exponential (SE) and Periodic (Pe). The NLML of the GP when applied to the full capacity vs.\ cycle data was used to evaluate the kernel combinations. The ranking of these results is shown in the bar-plot of Fig.~\ref{fig:Figure_gpKernelSearch}. This plot shows that four combinations achieve an NLML between 527 and 530, and hence perform similarly well. The performance then drops off more rapidly for the subsequent 5 combinations, and finally the last combination (Pe+Pe, NLML = 97.1) performs significantly worse than the others. This is perhaps not surprising given the lack of any exactly periodic structure in the data, let alone a superposition of two periodic components.
	
	\head{Kernel function selection}
	Although Ma3+Ma3 was the highest ranked pair, it performed only marginally better than the second ranked pair, Ma5+Ma3 (red bar in Fig.~\ref{fig:Figure_gpKernelSearch}). Hence we chose the latter for subsequent analysis because the contributions of each base kernel are easily \revB{interpretable.}

	\subsection{Kernel function decomposition}
\begin{figure*}[hbt]
		\centering
		\includegraphics[width=0.98\textwidth]{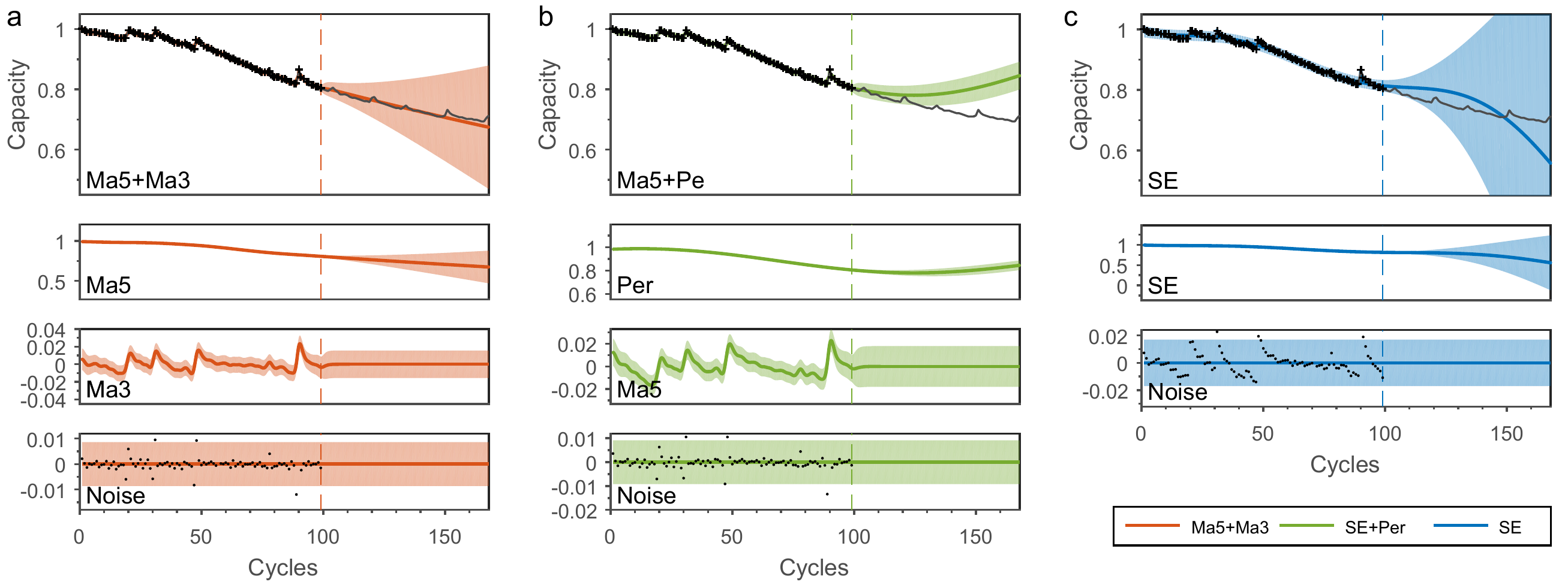}
		\caption{\caphead{Decomposition of posterior functions into constituent kernels.} The top plot in each column shows the posterior mean and \revB{credibility interval} of the compound kernel. For this and all subsequent figures, the black markers indicate data-points used for training, and the continuous black line indicates testing data, the coloured line indicates the mean of the posterior, and the correspondingly coloured shaded region indicates the area enclosed by the mean plus or minus \revB{two standard deviations}. The vertical line indicates the extent of the training data. \textbf{a}, Ma5+Ma3 (the selected kernel pair); \textbf{b}, SE+Per (a poorly performing kernel pair); \textbf{c}, SE (a singleton kernel)}
		\label{fig:Figure_gpDecomposeCov}
	\end{figure*}

	\head{Covariance decomposition}
	In order to highlight the significance of kernel function selection, the posterior mean and covariance for selected kernel pairs are decomposed into their constituent contributions in Fig.~\ref{fig:Figure_gpDecomposeCov} (using equations 2.17 and 2.18 from \cite{duvenaud2014automatic}).
	
	\head{Part A:} Fig.~\ref{fig:Figure_gpDecomposeCov}a shows the decomposition of the selected Ma5+Ma3 kernel pair, evaluated using 55\% of the cycle-capacity data and tested on the remainder. The sub-plots beneath each main plot show the individual contributions, including the noise covariance (which is implicitly included in each model). It can be seen that the Ma5 term captures the smooth long-term downward trend as desired, with an increasing uncertainty as it is projected into the future; the Ma3 term captures the short term variation; and the \revB{noise} term models the remaining small scale variation in the data. As a result, the extrapolation performance at this particular cycle is quite good, as indicated by the close match between the mean prediction and the true data for the remaining cycles.
	\head{Part B:} Fig.~\ref{fig:Figure_gpDecomposeCov}b shows the result for a kernel pair which performed less well (NLML = 500, Fig.~\ref{fig:Figure_gpKernelSearch}). In this case, the long-term trend is captured by the periodic component, whilst the \revC{Ma5} term is forced to model the short term variation. This indicates that there is little actual periodic structure present in the data, since the optimised length-scale of the periodic term is similar to the time-scale of the data, and hence only half a cycle of the periodic term is modelled. The predictions from this model indicate that the capacity will increase in the subsequent 100 cycles, before decreasing again and then repeating this behaviour periodically, which is clearly unrealistic.
	\head{Part C:} Finally, Fig.~\ref{fig:Figure_gpDecomposeCov}c shows the decomposition of a singleton SE kernel function (i.e.\ not the sum of two base kernels). In this case, the SE term is forced to try to model both the long and short term trends in the data. This results in the long-term trends being heavily influenced by the short term variations. Hence, when the short term step increase at $\sim$ cycle 90 is reached, the model predicts a smooth increase (then subsequent decrease) in the gradient, which is unrepresentative of the data. Moreover, there is obvious structure still present in the \revB{noise} contribution (bottom sub-plot), which indicates that not all of the structure in the data has been captured by the model. Both of these attributes are clearly undesirable and lead to poor extrapolation performance.

	\subsection{Short-term lookahead prediction}
\begin{figure*}[hbt]
		\centering
		\includegraphics[width=0.98\textwidth]{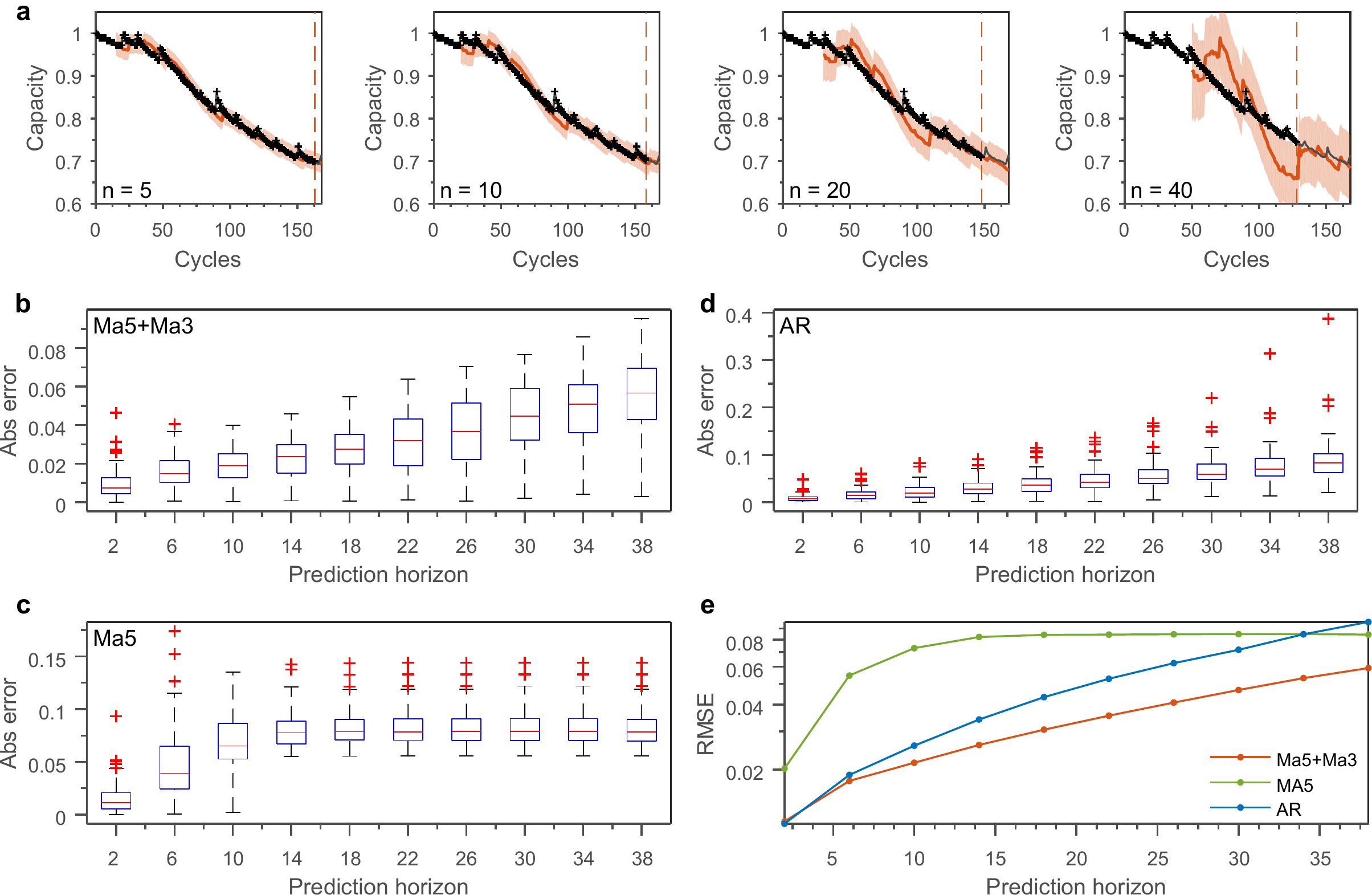}
		\caption{\caphead{Short term lookahead prediction performance.} \textbf{a}, Posterior \revB{distribution} of the Ma5+Ma3 GP for a range of different prediction horizons as indicated. \textbf{b-d}, Box-plot of capacity prediction errors for \textbf{b}, the Ma5+Ma3 GP, \textbf{c}, the singleton Ma5 GP and \textbf{d}, an autoregressive moving average of order 10. \textbf{e}, $\text{RMSE}_Q$ of each method plotted on the same axis with the y-axis in log scale for ease of comparison.}
		\label{fig:Figure_gpLookahead1}
	\end{figure*}

	Having selected a suitable kernel function, we now investigate \revC{the} extrapolation performance using training data up to various different current cycles, $c$. Fig.~\ref{fig:Figure_gpLookahead1} shows the performance of the method for n-step lookahead forecasting. Fig.~\ref{fig:Figure_gpLookahead1}a shows the posterior mean and covariance using prediction horizons of 5, 10, 20 and 40 cycles. For each cycle number, the posterior is obtained using data up to the current cycle, and the mean and standard deviation \revC{are} evaluated at the cycle n steps ahead of the current cycle. This is repeated at every cycle up until the vertical dashed line (equal to the number of the last cycle minus the size of the prediction horizon). Hence, the posterior mean and variance shown in the plots is the amalgamated posterior from all of these cycles.
	
	\head{Performance}
	The plots show that the method is highly accurate for relatively small $n$ but that the performance diminishes as $n$ is increased.
	This is hardly surprising given that we have no a-priori reason to believe that the capacity data up to a given cycle has strong predictive capabilities for distant future cycles.
	
	\head{Value of kernel selection}
	However, it is clear that the principled selection of the kernel function has been advantageous, since the method clearly outperforms the singleton Ma5 GP. Fig.~\ref{fig:Figure_gpLookahead1}b shows box-plots of the extrapolation error for various prediction horizons for the Ma5-Ma3 GP, whilst Fig.~\ref{fig:Figure_gpLookahead1}c shows the same data for the singleton Ma5 GP.
	Lastly, we evaluated a more conventional time-series approach, an autoregressive moving average of order 10 (i.e. using the 10 most recent data-points for training at each cycle), which is shown in Fig.~\ref{fig:Figure_gpLookahead1}c.
	
	The Ma5+Ma3 GP has the best performance of these three, as shown by \rev{Fig.~\ref{fig:Figure_gpLookahead1}e} which plots the RMSE against prediction horizon of each of the three cases on a single axis for ease of comparison.
	This can be attributed to the fact that the additive kernels are capable of handling the processes of different scales\revA{ -- }with the short term variation being handled by one of the constituent kernels and the long-term downward trend by the other.

	\subsection{Remaining useful life prediction}
	
	\head{Prediction requirements}
	Depending on the requirements of the system, predicting future capacity 10 cycles ahead with reasonable accuracy  may be sufficient to facilitate corrective action. In some cases, however, it may be desirable to accurately estimate the \revC{remaining useful life} (RUL) from the earliest stage possible, in which case accurate long-term prediction is important.

	\begin{figure*}
		\centering
		\includegraphics[width=0.98\linewidth]{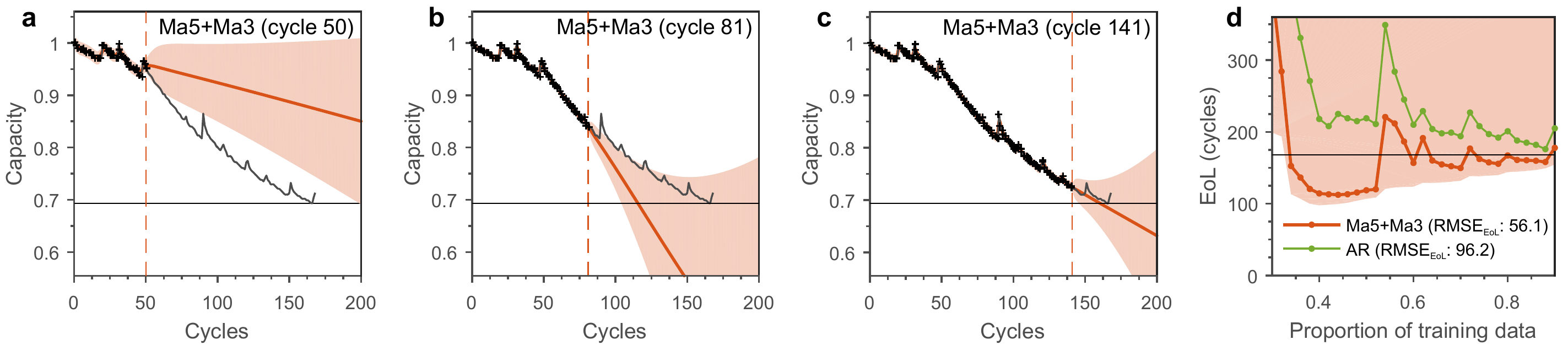}
		\caption{\caphead{End of life (EoL) estimation performance.} \revB{\textbf{a-c}}, Posterior distribution of the Ma5+Ma3 GP at 3 different proportions of the training data. The horizontal black line indicates the capacity at the EoL. \textbf{d}, Predicted cycle no.\ at EoL vs.\ training data proportion for the GP and AR methods. The shaded region indicates the confidence interval in the EoL prediction. Note that there is no green shaded area since the AR does not provide confidence bounds. The horizontal black line indicates the true EoL.}
		\label{fig:Figure_gpNaiveRuL}
	\end{figure*}

	\head{EoL performance}
	Fig.~\ref{fig:Figure_gpNaiveRuL} shows the performance of the method for estimating the \revC{end of life} (EoL) at three different current cycle values. The method shows some desirable properties, such as converging to the correct EoL estimate as more training \revC{data are} acquired and having large credible intervals when the extrapolation is far into the future. It also outperforms the baseline autoregressive model.
	However, the EoL predictions are poor at the initial cycles when there \revC{are} limited training data available. For instance, it can be seen from 
	\revB{Fig.~\ref{fig:Figure_gpNaiveRuL}a-b}
	that the EoL is first severely over-estimated and then severely underestimated as new \revC{data are} received. This results in only moderate overall performance as indicated by the
	\revB{corresponding $\text{RMSE}_{\text{EoL}}$ values (see inset of Fig.~\ref{fig:Figure_gpNaiveRuL}d.}
	\revB{This plot also shows the credibility intervals in the EoL estimate. These were obtained by extrapolating the upper and lower confidence intervals in the capacity estimates until they reached the EoL value. In some cases the upper confidence interval never crosses the lower threshold and hence the upper EoL estimate is very large or infinite (extending beyond the upper limit of the y-axis in this plot). This is an unfortunate consequence of the fact that the model is not restricted to be monotonic, as we discuss in Section~\ref{sec-conclusions}.
	However, it is promising that from about a third of the training data onwards, the true EoL estimate always remains within the lower confidence interval.}
	

\section{Encoding exponential degradation via EMFs -- Results \label{sec:results-EMFs}}
	\head{Overview}
	In order to improve \revC{the }long term predictive forecasts, we now consider a single-output GP with an explicit mean function based on a battery degradation model from the literature.
	The dataset in this case consists of capacity vs.\ cycle data extracted from \cite{wang2013prognostics}  (see \revB{Appendix \ref{sec:appendix-data}}, and Fig.~\ref{fig:Figure_gpData}b). Here, we apply the method to Cell B3 since this exhibits the greatest deviation from the exact exponential decay behaviour of the model, and hence benefits most from the additional non-parametric contributions provided by the GP.
	
	We use an explicit mean function of the form $\revB{m(x)} = a_1 + a_2 \, \exp{(a_3 \, x)}$. The model parameters are thus given by $\theta_{\text{deg}} = [a_1, a_2, a_3]$. This function is equivalent to the degradation model used by Goebel et.\ al~\cite{goebel2008prognostics}. It could also be viewed as a special case of the three-parameter degradation model used by Wang et al. \cite{wang2013prognostics} with the \revB{``empirical factor''} set to zero (i.e.\ $g = 0$ in Eq.~(21) of \cite{wang2013prognostics}).
	
	We consider three different GP models:
	\begin{enumerate}
		\item[a.] $y = f(x) + \revB{m(x)}$, where $f(x) \sim GP(0,  \kappa_{\text{Ma3}})$
		\item[b.] $y = f(x) + \revB{m(x)}$, where $f(x) = \revB{\varepsilon} \sim GP(0,  \kappa_{\text{noise}})$
		\item[c.] $y \sim GP(0,  \kappa_{\text{Ma5} + \text{Ma3}})$
	\end{enumerate}
	
	Model (a) assumes that the response consists of the specified exponential mean function plus a GP with Ma3 covariance. Model (b) is identical to (a) but with a noise covariance; since the covariance is simply white noise, this is essentially just a parametric model. Model (c) is the basic GP that gave best results in the previous section (i.e.\ with a \revC{zero} mean function but with a covariance function consisting of a sum of Ma5 and Ma3 terms).
	
\begin{figure*}[hbt]
		\centering
		\includegraphics[width=0.98\textwidth]{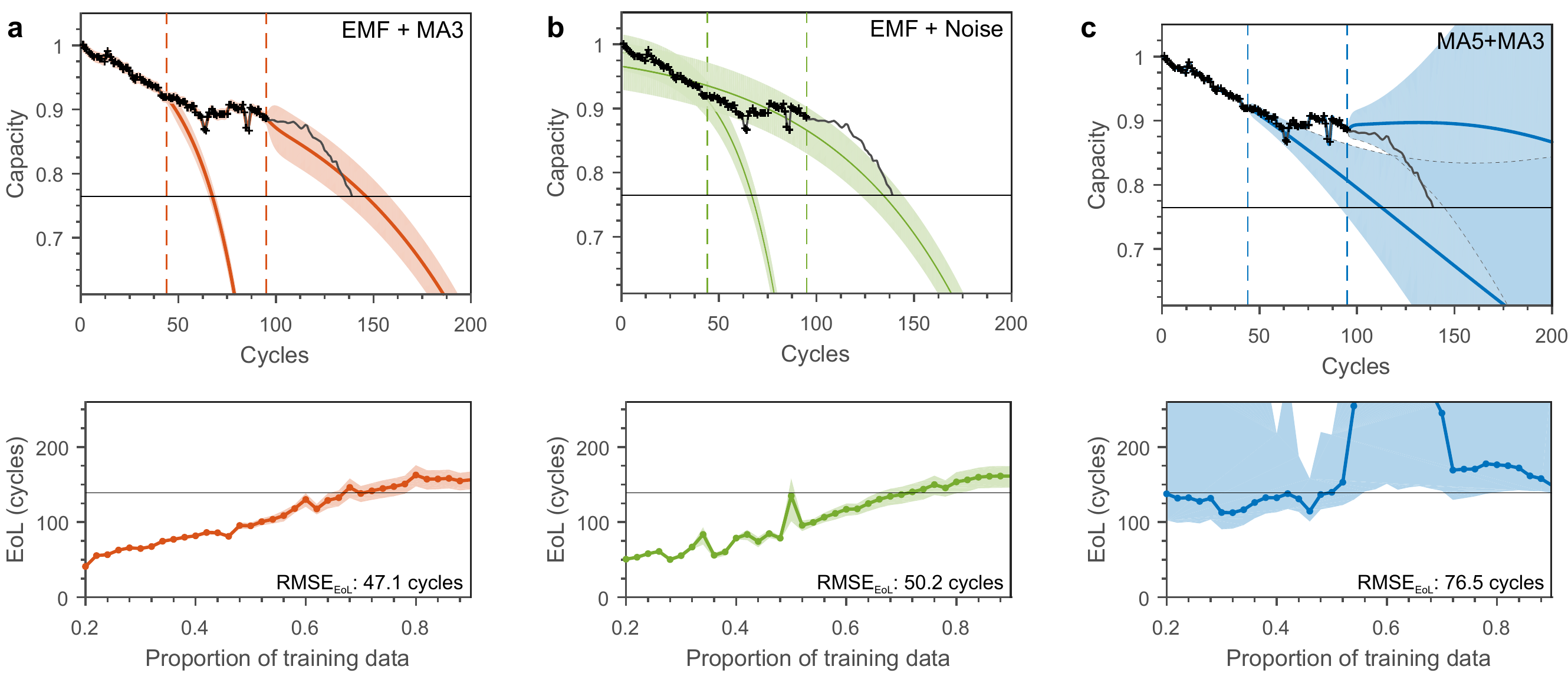}
		\caption{\caphead{Explicit mean function GP results.} \textbf{a-c},  (top) \revB{Posterior} distribution at 2 different proportions of the training data for \textbf{a}, EMF+Ma3, \textbf{b}, EMF+Noise and \textbf{c}, Ma5+Ma3. The horizontal black line indicates the capacity at the EoL. (bottom) Predicted cycle no.\ at EoL vs.\ training data proportion for the corresponding methods. The shaded region indicates the credible interval in the EoL prediction. The horizontal black line indicates the true EoL cycle.
		}
		\label{fig:Figure_gpExplicitMean}
	\end{figure*}

	Fig.~\ref{fig:Figure_gpExplicitMean} shows the posterior predictions at two arbitrarily chosen cycles (top) and the estimated EoL (bottom) for each of the three cases.

	It can be seen from this figure that model (c) performs poorly in the region of $\sim$60 cycles, where the capacity temporarily levels off. This is because the model has no prior assumption encoded about the degradation behaviour (other than the smoothness assumptions encoded by the covariance) and hence, when only the data up until this time step \revC{are} used for training, it predicts a subsequent upwards trend in the capacity. This can be seen from the bottom plot of Fig.~\ref{fig:Figure_gpExplicitMean}c in the same region, which shows that the EoL prediction extends beyond the upper limit of the y-axis.
	In contrast, models \revC{(a) and (b)} cope with this temporary stationary behaviour and correctly predict a continuing exponential degradation for subsequent time steps.
	As a result of this, models (a) and (b) also have lower overall \revB{$\text{RMSE}_{\text{EoL}}$} values than model (c).
 	Hence, these results show that the use of an explicit mean function improves the overall accuracy of the EoL predictions.
 	
 	Comparing model (a) against model (b), we can see the advantage of using a GP model over a purely parametric model.
 	Since model (b) assumes that any deviation from the exponential model must be noise (since the kernel function is defined to be white noise), the optimised noise parameter becomes quite large and the credible interval increases to encompass the spread in observed capacities. In contrast, model (a) models the deviations with an Ma3 kernel, and hence can fit the non-exponential trend quite well, whilst maintaining a sensible assumption for the noise levels.
	
	Lastly, it should be noted that models (a) and (b) are both overconfident in their predictions (the \revB{credibility intervals in} the EoL estimation in the bottom plots are too narrow and do not encompass the true EoL for most of the training proportions).
	Moreover, in some cases the uncertainty is seen to decrease with increasing cycle number, in particular in model (b). This is obviously not desirable behaviour and is an indication that the assumption of the functional form of the underlying model (i.e. the exponential degradation model) is not entirely valid for this data.

\section{Capturing cell-to-cell correlations via multi-output GPs -- Results\label{sec:results-multi-output}}

	The final example we consider is a multi-output GP.
	The dataset in this case consists of capacity vs.\ time
	data from three cells with randomised \revC{load profiles} (see \revB{\ref{sec:appendix-data}}, and Fig.~\ref{fig:Figure_gpData}c).
	Since the cell cycling is randomised, a parametric degradation model (which is a function of the number of cycles) would be unsuitable. Hence, we rely on exploiting data from existing cells to improve the long-term forecast.
	We apply the method to cell C3 using data from \revC{cells }C1 and/or C2 for training as follows.
	
	We considered four different models:
	\begin{enumerate}
		\item[a.] A multi-output GP with 3-outputs: the capacity data for cells C1-C3. The model is trained on \emph{all} capacity data for cells C1 and C2, along \revB{with} data up until the current cycle for cell C3, in the same manner as the previous test cases.
		\item[b.] A multi-output GP with 2-outputs: the capacity data for cells C1 and C3 (omitting C2). The model is trained on \emph{all} of the capacity data for cell C1, along with data up until the current cycle for cell C3 as before.
		\item[c.] A multi-output GP with 2-outputs as in (b) but provided with data from C2 instead of C1.
		\item[d.] A standard single-output GP trained using just data up to the current cycle for C3.
	\end{enumerate}
	
	The multi-output models were defined as described in Section~\ref{sec:methods-multi-output}, with the $\kappa_{\text{MOGP}} = \kappa_l \times \kappa_x$ where $\kappa_l$ is the label covariance matrix (Eq.~\ref{eq:spherical-parametrisation}), and $\kappa_x = \kappa_{\text{Ma5}+\text{Ma3}}$.
	The covariance of the single output model was simply $\kappa = \kappa_x = \kappa_{\text{Ma5}+\text{Ma3}}$.

\begin{figure*}[!hbt]
		\centering
		\includegraphics[width=0.98\textwidth]{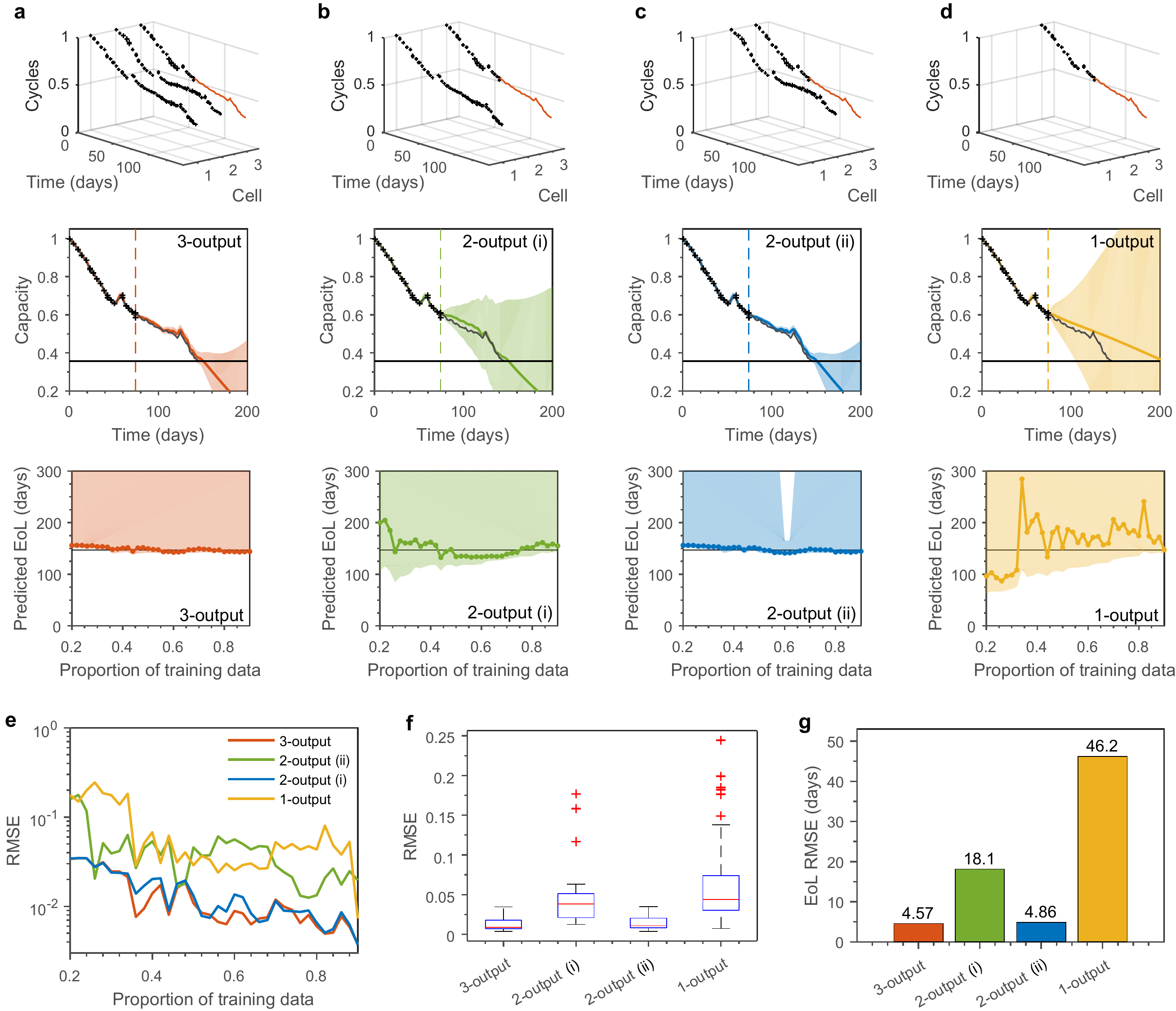}
		\caption{\caphead{Multi-output GP results.} \textbf{a-d}, Data and results for \textbf{a}, 3-output, \textbf{b}, 2-output (i), \textbf{c} 2-output (ii), and \textbf{d} single-output GPs. (top) Battery data; black markers indicate training data and red line indicates unseen testing data. (middle) Posterior distribution using 40\% of the training data and (bottom) Predicted cycle no.\ at EoL vs.\ training data proportion.
		\textbf{e}, RMSE vs.\ proportion of training data; \textbf{f}, Box-plot of RMSE data; and \textbf{g}, Bar plot of RMSE on EoL prediction.
		Note that the colours of the lines/shaded regions in \textbf{a-d} correspond to the colours of the lines and bars in \textbf{e} and \textbf{g}.}
		\label{fig:Figure_gpMulti1}
	\end{figure*}

	The training data and results for these four cases are shown in Fig.~\ref{fig:Figure_gpMulti1}.
	Figs.~\ref{fig:Figure_gpMulti1}a-d show the data used for training at a selected time step (top), the posterior predictions at the corresponding time step (middle), and the estimated EoL as a function of the number of cycles of \revC{cell C3} training data used (bottom) for each of the four cases.
	Figs.~\ref{fig:Figure_gpMulti1}e-g show summary statistics for all four cases: Fig.~\ref{fig:Figure_gpMulti1}e shows $\text{RMSE}_Q$ as a function of the proportion of the training data (for cell C3) used.
	Fig.~\ref{fig:Figure_gpMulti1}f shows a boxplot of the same data.
	Fig.~\ref{fig:Figure_gpMulti1}g shows a bar plot of $\text{RMSE}_{\text{EoL}}$.
	
	It can be seen from these results that the multi-output models (a-c) have better predictive performance than the single output model (d). For instance, in the middle subplots in Fig.~\ref{fig:Figure_gpMulti1}a-d (which show the posterior prediction at 75 days), all of the multi-output models accurately track the future capacity up until the predicted EoL, whereas the single output method does not anticipate the sudden drop in capacity at $\sim$125 days, and hence over-predicts the subsequent capacity. Indeed, the estimates of the EoL of \revC{cell C3} are quite accurate throughout the entire range of training data (bottom subplots) for the multi-ouput methods, but are shown to fluctuate significantly as additional training \revC{data are} received in the single-output case.
	This is also reflected in the overall $\text{RMSE}_Q$ and $\text{RMSE}_{\text{EoL}}$ values depicted in Figs.~\ref{fig:Figure_gpMulti1}e-g.

	Interestingly, there are significant differences in the performance of models b and \rev{c} (which are both two output models); for instance, model b, which is trained on \revC{cell~C1}, has an $\text{RMSE}_{\text{EoL}}$ of 18.1, whereas model c, which is trained on \revC{cell~C2}, has an $\text{RMSE}_{\text{EoL}}$ of 4.86. Likewise, there is greater uncertainty (larger error bounds) in the posterior predictions of model b than those of model c (see e.g. middle subplots of Fig.\ref{fig:Figure_gpMulti1}b and c).
	This indicates that the capacity trend of \revC{cell C3} shares a stronger correlation with that of \revC{cell C2} than that of \revC{cell C1}.
	
	It is perhaps even more interesting to note that model~a (which has 3-outputs) is not negatively affected by the inclusion of data from Cell 1. Rather, the model naturally puts more weight on the data from \revC{cell C2} (than from \revC{cell C1}) since this shows a stronger correlation with \revC{cell C3}. As a result, the performance of model a turns out to be superior to either model b or model c in this case (Figs.~\ref{fig:Figure_gpMulti1}f-g).
	
	There are a number of obvious caveats to be aware of in the present case: the method has performed particularly well here because the dataset consists of cells cycled under the same thermal conditions with statistically equivalent applied current profiles\footnote{Note that the applied current profiles are generated by a stochastic algorithm\revC{~\cite{bole2014randomized}} and hence they are \emph{not} identical from cell to cell. However, since the algorithm is the same for each cell, the global properties of these current profiles are the same, and so they manifest similar degradation behaviour.}. Hence, there were strong correlations between the data for these cells. Whilst this could occur in practice in some limited cases (for instance, multiple cells in a pack could experience the same conditions), in most cases of interest we would wish to use data from identical cells but cycled under differing conditions.
	The method could of course be extended to account for such differences by including additional inputs, such as temperature or depth of discharge, although this would require data from many more cells to cover a broad range of operating conditions.
	For such (larger) datasets, more efficient implementation methods must be used, such as those discussed in Section~\ref{sec:methods-multi-output}.

	However, the present results indicate that the multi-output method has promise.	
	In particular, it is favourable over previous GP capacity estimation methods that use data from identical cells, which merely identify an optimal prior estimate for the parameters of a parametric model, which are then updated \revB{sequentially}~\cite{liu2013prognostics}. In such a setting the advantage of the prior information is quickly lost. In contrast, the present method exploits correlations in cell behaviour over the duration of the cell life, which leads to the improved results obtained here.
	
\section{Conclusions\label{sec-conclusions}}

This paper has demonstrated the applicability of GPs to battery capacity forecasting, and highlighted some of their key advantages for this application.
We have shown the advantage of using compound kernel functions for capturing complex behaviour and highlighted the importance of proper kernel function selection by means of optimising w.r.t.\ the NLML.
We have shown that using an explicit mean function based on known degradation models, it is possible to improve the predictive performance of the baseline GP.
Indeed, we argue that one should never just consider a purely \revB{data-driven} approach if prior information on the functional form of the underlying model is available; or a purely parametric approach if the model information is not known with certainty, since in either case one would be neglecting a valuable source of information.
Lastly, we have shown that multi-output models can effectively exploit data from multiple cells to significantly improve forecasting performance.
The main bottleneck of this approach is the computational cost of handling large numbers of outputs; although we applied this method to a small dataset of just 3 cells in the present case, efficient approaches exist which allow scaling to very large numbers of outputs.
	
The present work aims to highlight just some of the advantages of GPs. Several extensions/variations on this work are also possible. For instance:

\begin{itemize}
	\item We know a-priori that the normalised cell capacity must take a value between 0 and 1. In the present work, this is not enforced and at times the GP may predict values outside of this range (in particular for single-output GPs with limited training data). One way to enforce positive values is to apply a \revB{logit}-transform to the data as a pre-processing step and apply a GP to the resulting data, then reverse transform the result.
	\item The notion that the future capacity depends only on the past values is na\"ive. A more comprehensive study could include DoD, temperature, idle time etc.\ as inputs. Whilst such dependencies have been considered in a parametric framework in previous works (e.g.~\cite{wang2011cycle}), the application in a GP framework has not yet been considered, Such an approach would require data from many more cells to cover a broad range of operating conditions.
	\item \rev{Cells often exhibit regime changes during aging (e.g.\ a transition from a linear capacity vs.\ cycle trend to a non-linear regime with accelerated aging~\cite{you2017diagnosis}). GPs have the capability to account for such transitions; specifically, \emph{change-point kernels}~\cite{saatcci2010gaussian,garnett_sequential_2010} could be used to locate change points in an online manner, and hence more accurately model the regimes before and after a transition than would be possible with a single model over the entire domain.}
	\item The GP framework could also be applied using higher dimensional input data, such as open circuit voltage curves or electrochemical impedance spectra acquired at periodic cycles.
\end{itemize}

We hope that the present study has provided motivation for further study of GPs applied to battery capacity forecasting.

\FloatBarrier

\appendix
	
	\section{Data\label{sec:appendix-data}}
	The datasets (Fig.~1) consist of capacity vs.\ cycle data obtained from either open-access NASA repositories or extracted directly from the plots in previous papers. In each of these cases, the results obtained from a single selected cell were presented in this paper; similar results were obtained for the other cells in each dataset, but for brevity these were not presented.

	\head{Dataset A}
	Dataset A is obtained from the NASA Ames Prognostics Center of Excellence  \hyperref{https://ti.arc.nasa.gov/tech/dash/pcoe/prognostic-data-repository/}{category}{name}{Battery Data Set}\cite{saha2007battery}.
	The experiments consisted of applying several charge-discharge cycles to a number of commercially available 18650 lithium-ion cells at room temperature in order to achieve accelerated aging.
	Charging was carried out via a constant-current, constant-voltage regime: charging at constant current at 1.5~A until the voltage reached the cell upper voltage limit of 4.2~V, then applying a constant voltage until the the current dropped to 20~mA.
	Discharging was carried out at a constant current of 2 A until the cell voltage fell to 2.7~V, 2.5~V, 2.2~V, and 2.5~V for batteries 5, 6, 7, and 18, respectively.
	The experiments were stopped when the batteries had lost 30\% of the \revC{initial} capacity.
	Additional data (including temperature and electrochemical impedance) are also provided in the online repository, although these were not used in the present study.
	Full details of the experiments are available in Ref.~\cite{saha2008uncertainty}.
	
	Batteries  5, 6, 7, and 18 (in the numbering of the online repository) were chosen to be analysed in the present work, since these have the most data-points; and because they have previously been chosen for analysis in earlier works~\cite{he2011prognostics,liu2013prognostics}, and hence the present selection facilitates a comparison with those works.
	For consistency, these cells are labelled as A1, A2, A3 and A4 respectively in the present paper (Fig.~\ref{fig:Figure_gpData}a). The results for Cell~A1 are presented in Section~\ref{sec:basic-gp}.
	
	\head{Dataset B} Dataset B was obtained by manually extracting the data from the capacity vs.\ cycle plots in Fig.~1 of Ref.~\cite{he2011prognostics}, using Matlab GRABIT (a tool for extracting raw data from plot images). These data were originally obtained from charge-discharge experiments on a 0.9~Ah lithium-ion cell. The discharge rate was 0.45~A and the currents were cut off at the upper and lower voltage limits specified by the battery manufacturer.
	Further details of the experiments are available in Ref.~\cite{he2011prognostics}.
	These cells are denoted B1, B2, B3 and B4 in the present paper (Fig.~\ref{fig:Figure_gpData}b). The results for Cell~B3 are presented in Section~\ref{sec:results-EMFs}.
	
	\head{Dataset C}
	Dataset C is obtained from the NASA Ames Prognostics Center of Excellence Randomized Battery Usage
	\href{https://ti.arc.nasa.gov/tech/dash/pcoe/prognostic-data-repository/}{Data Set} \cite{bole2014randomized}.
	The experiments consisted of applying randomized sequences of current loads	ranging from 0.5~A to 4~A to a number of  LG Chem. 18650 lithium-ion cells at a range of environmentally controlled temperatures in order to achieve accelerated aging.
	The sequence was randomized in order to better represent practical battery usage. After every fifty randomized discharging cycles, the capacity was measured via a low-rate charge-discharge cycle, and the electrochemical impedance were measured via an EIS sweep.
	The experiments are described in detail in Ref.~\cite{bole2014adaptation}.

	The first three cells from Dataset 1, denoted RW9, RW10 and RW11 (in the numbering and labelling of the online repository) were chosen for analysis. These cells were cycled at room temperature and express highly non-linear and non-parametric behaviour. The cells are labelled as C1, C2 and C3 in the present paper (Fig.~\ref{fig:Figure_gpData}c). The results for Cell~C1 are presented in Section~\ref{sec:results-multi-output}.

\FloatBarrier
\bibliographystyle{elsarticle-num}
\bibliography{GP_Paper}

\section*{Acknowledgments}
This work was funded by an RCUK Engineering and Physical Sciences Research Council grant, ref. EP/K002252/1.

\end{document}